\documentclass{article}
\usepackage{times}
\usepackage{xcolor}
\usepackage{booktabs}
\usepackage{colortbl}  
\usepackage{amsfonts}
\usepackage{latexsym,amssymb,amsmath,amsfonts} 
\usepackage{graphicx}
\begin{document}
\noindent
{\Large LANDAUER ENTROPY OF SPACETIME}

\vskip.5cm
\noindent
{\bf  J.M. Isidro}$^a$, {\bf B. Koch}$^{bc}$ and {\bf \'A. Rinc\'on}$^{ade}$\\
$^a$ Instituto Universitario de Matem\'atica Pura y Aplicada,\\ Universitat Polit\`ecnica de Val\`encia, Valencia 46022, Spain\\
{\tt joissan@mat.upv.es}\\
$^b$ Institut f{\"u}r Theoretische Physik, \\ Technische Universit{\"a}t Wien, Wiedner Hauptstrasse 8–10, A-1040 Vienna, Austria\\
$^c$ Pontificia Universidad Cat{\'o}lica de Chile
Instituto de F{\'i}sica, Pontificia Universidad Cat{\'o}lica de Chile,
Casilla 306, Santiago, Chile\\
{\tt benjamin.koch@tuwien.ac.at}\\
$^d$ Departamento de F{\'i}sica, Universidad del B{\'i}o-B{\'i}o,
Casilla 5-C, Concepci{\'o}n, Chile.\\
$^e$ Research Centre for Theoretical Physics and Astrophysics, Institute of Physics, Silesian University in Opava, Bezručovo náměstí 13, CZ-74601 Opava, Czech Republic\\
{\tt angel.rincon@physics.slu.cz}\\
\vskip0.5cm
\noindent
\today
\vskip0.5cm
\noindent
{\bf Abstract} Based on Landauer's principle, we provide a geometrical definition for the entropy of a given static, spherically symmetric spacetime. Considering a congruence of geodesics across a surface, one defines the entropy of a congruence as the surface integral of the entropy of the constituent geodesics. Under certain mild assumptions, we establish a second law for the entropy function thus defined (Landauer entropy), and relate it to Bekenstein--Hawking entropy.

\tableofcontents

\section{Introduction}

In the context of gravitational physics, the concept of entropy plays a central role. Entropy is conventionally defined via the well--known Boltzmann expression $S=k_B\ln(W)$, where $k_B$ is the Boltzmann constant, and $W$ denotes the number of microscopic configurations compatible with a given macroscopic state. This definition, which comes from a statistical treatment, has a remarkable impact in the context of black hole physics, where the Bekenstein–Hawking entropy links the entropy of a black hole with its corresponding area $A$ evaluated on its event horizon: $S=A/(4G\hbar)$. At this point it becomes clear that this link opens a new avenue between thermodynamics and gravity.

From a purely thermodynamic point of view, the entropy is  indispensable to a full description of the system. This latter assertion is true because such a concept can be trivially interpreted as a measure of disorder and irreversibility of a system, and through the second law of thermodynamics, it is responsible for the direction of spontaneous evolution (in isolated systems).
Entropy plays a crucial role in different ways, for instance:
\begin{itemize}
    \item[{\it i)}] in determining the stability of self-gravitating objects;
    \item[{\it ii)}]in shaping the thermodynamic evolution of the universe, and
    \item[{\it iii)}]in establishing fundamental bounds on the information content that can be stored within a finite region of spacetime.
\end{itemize}

A pivotal connection between information theory and thermodynamics is established by Landauer’s principle \cite{LANDAUER}. This principle asserts that the logical erasure of one bit of information in a computational process is necessarily accompanied by a minimum heat dissipation to the environment of $k_B T \ln 2$,  where $T$ is the temperature of the thermal reservoir. As a consequence, information erasure produces a corresponding increase in the physical entropy of the surroundings by at least $k_B \ln 2$. 

This paper is organised as follows.
In section \ref{entrograv} we relate Landauer's principle to the notions of entropic gravity advocated in ref. \cite{VERLINDE}.
In section \ref{Lan} we formally introduce the Landauer entropy associated with a given metric and establish its monotonicity. In section \ref{results} we present our main results in a series of tables for clarity, while the computational details are deferred to an appendix. In section \ref{Disc} we summarise and discuss our results. We use the signature $\{-,+,+,+\}$.

\section{Landauer's principle and entropic gravity}\label{entrograv}

Landauer's principle \cite{LANDAUER} is the assertion that the erasure of 1 bit of information at temperature $T$ costs an energy $\Delta E$ that is bounded from below by the value $k_BT\ln 2$:
\begin{equation}
\Delta E\geq k_BT\ln 2.
\label{1}
\end{equation}
Division by $T$ leads to an inequality that is strongly reminiscent of the second law of thermodynamics:
\begin{equation}
\Delta S\geq k_B\ln 2,
\label{2}
\end{equation}
where $\Delta S = \Delta E/T$.
The quantity $k_B\ln 2$ appears to be a {\it quantum of entropy}\/, similarly to the quantum of action represented by Planck's constant $\hbar$. 

It is well known that the presence of a weak gravitational field described by the Newtonian potential $\phi$ modifies the Minkowski metric to
\begin{equation}
{\rm d}s^2=-\left(1+2\frac{\phi}{c^2}\right)c^2{\rm d}t^2+\left(1+2\frac{\phi}{c^2}\right)^{-1}{\rm d}r^2+r^2{\rm d}\Omega_2^2,
\label{dos}
\end{equation}
where ${\rm d}\Omega_2^2$ is the standard round metric on the 2--sphere $\mathbb{S}^2$. 
Any statistical ensemble immersed in such a background will be sensitive to the presence of the gravitational potential $\phi$.
In particular, energy differences will acquire a 
factor $1+\phi/c^2$, due to 
Tolman's effect~\cite{TOLMAN}. Correspondingly, Landauer's principle (\ref{1}) gets replaced with \cite{HERRERA}
\begin{equation}
\Delta E\geq k_BT\left(1+\frac{\phi}{c^2}\right)\ln 2.
\label{3}
\end{equation}
Dividing by $T$ we get an entropic--like inequality
\begin{equation}
\Delta S\geq k_B\left(1+\frac{\phi}{c^2}\right)\ln 2.
\label{4}
\end{equation}
Now the factor $1+\phi/c^2$ is the weak field limit of $\sqrt{\vert g_{tt}\vert}$:
\begin{equation}
\sqrt{\vert g_{tt}\vert}=\sqrt{1+2\frac{\phi}{c^2}}=1+\frac{\phi}{c^2}+\ldots 
\label{5}
\end{equation}
This suggests defining a Landauer entropy $S$ associated with an arbitrary Lorentzian metric $g_{\mu\nu}$ as
\begin{equation}
S=k_B\sqrt{\vert g_{tt}\vert},
\label{6}
\end{equation}
after absorbing the factor $\ln 2$ within $k_B$.

In a related venue, {\it gravity as an entropic force}\/ is the notion that the gravitational interaction might be a coarse--grained, macroscopic phenomenon, arising as the effective theory of some underlying degrees of freedom. Thus, {\it e.g.}\/ in ref \cite{VERLINDE}, the Newtonian force as given by the gradient of a potential function, ${\bf F}=-\nabla\phi$,\footnote{Here ${\bf F}$ denotes force per unit mass.}  is replaced by the gradient of an entropy funcion, ${\bf F}=K\nabla S$, where the constant $K$ will correct dimensions. The proposal that gravity cannot be a fundamental force has been around for some time \cite{JACOBSON1, PADDY1, SINGH}, especially after the realisation that spacetime exhibits thermodynamical properties. 

Beyond black holes, singularities and horizons, these arguments support the view that one can assign an entropy (that we will call Landauer entropy) to any given gravitational potential $\phi$ and, more generally, to any given spacetime metric. In this letter we will explore how to define a Landauer entropy on a given spacetime, using solely macroscopic arguments, {\it i.e.}\/, without resorting to assumptions as to the microscopic physics of spacetime.
 
Let us first recall the Newtonian limit, when all gravitational information is encoded within the gravitational potential $\phi$. Consider the motion of a test particle within the force field $-\nabla\phi$ leading to the potential increase $\delta\phi$. According to entropic gravity, this $\delta\phi$ produces an increase $\delta S$ in the entropy of space. These two increments carry opposite signs and are proportional to each other \cite{VERLINDE}:
\begin{equation}
\frac{\delta S}{k_B}\propto -\frac{\delta\phi}{c^2}.
\label{uno}
\end{equation}
Eq. (\ref{uno}) is the statement that, in the weak--field limit, equipotential surfaces are isoentropic surfaces: this defines an entropy function
\begin{equation}
S=-\frac{k_B}{c^2}\phi,
\label{10}
\end{equation}
in terms of the given Newtonian potential.\footnote{We will henceforth set $c=1$, $k_B=1$.} 

In what follows we will extend this notion of entropy beyond the Newtonian regime. For simplicity, the Lorentzian manifolds analysed here will be $4$--dimensional, spherically symmetric and static: 
\begin{equation}
{\rm d}s^2=-f^2(r){\rm d}t^2+\frac{{\rm d}r^2}{f^2(r)}+r^2{\rm d}\Omega_{2}^2.
\label{7}
\end{equation}
Above, $r$ is a radial coordinate and ${\rm d}\Omega_{2}^2$ is the round metric on the unit sphere $\mathbb{S}^{2}$.

\section{Landauer entropy} \label{Lan}

Inspired by Eq. (\ref{6}) we can define a Landauer entropy function depending only on the radial coordinate $r$:
\begin{equation}
S=\sqrt{\vert g_{tt}\vert}=f(r)=\frac{1}{\sqrt{g_{rr}}}.
\label{9c}
\end{equation}
By construction, $S$ is dimensionless. However, does it qualify as an entropy?

\subsection{Monotonicity}

The function (\ref{9c}) will qualify as an entropy whenever it satisfies a second law,
\begin{equation}
\frac{{\rm d}S}{{\rm d}t}\geq 0,
\label{seis}
\end{equation}
for all geodesic motions $x^{\mu}=x^{\mu}(t)$ of test particles with a fixed orientation. We have
\begin{equation}
\frac{{\rm d}S}{{\rm d} t}=\frac{\partial S}{\partial t}+\frac{\partial S}{\partial x^\mu}\frac{{\rm d}x^\mu}{{\rm d}t}
=\frac{\partial f}{\partial t}+\frac{{\rm d}x^\mu}{{\rm d}t}\frac{\partial f}{\partial x^\mu}
=\dot r(t)f'(r),
\label{doce}
\end{equation}
where $r=r(t)$ is any radial geodesic. Now $\dot r>0$ ($\dot r<0$) for outward (inward) directed motions, so the sign of (\ref{doce}) will be determined by that of $f'(r)$. 
Since $f'(r)$ is assumed continuous, it suffices to have $f'(r)\neq 0$ for all $r$ in order to ensure monotonicity of $S(t)$ in the variable $t$. If needed, exchanging inward with outward radial geodesics ($r(t)\rightarrow r(-t)$), or else flipping the sign in the definition of the entropy ($S(t)\rightarrow -S(t)$), will guarantee a monotonically {\it increasing}\/ behaviour of $S(t)$. Here we will adopt the latter criterion, in order to have {\it outgoing}\/ geodesics cause an entropy {\it increase}\/.

\subsection{A congruence of test particles}

So far we have defined a Landauer entropy $S(r)$ that is monotonically increasing along outgoing radial geodesics of test particles. We can consider a congruence of such geodesics traversing (not necessarily orthogonally) a 2--dimensional surface $\mathbb{S}$ (either open or closed). The Landauer entropy ${\cal S}$ of the congruence\footnote{Notice the different fonts ${\cal S}$ and $S$.} is defined to be proportional to the surface integral of the test--particle entropy $S$: 
\begin{equation}
{\cal S}\left[\mathbb{S}\right]=\frac{1}{4G\hbar}\int_{\mathbb{S}}{\rm d}A\, S=\frac{1}{4G\hbar}\int_{\mathbb{S}}{\rm d}A\,f.
\label{51c}
\end{equation}
Some remarks are in order.
\begin{itemize}
\item[{\it i)}] The notation ${\cal S}\left[\mathbb{S}\right]$ stresses the fact that (\ref{51c}) assigns a number ${\cal S}$ to a surface $\mathbb{S}$.

\item[{\it ii)}] In the particular case that the congruence traverses $\mathbb{S}$ orthogonally, the latter is defined by $r={\rm const}$.\footnote{Plus possibly some more requirements on $\theta$, $\varphi$, if $\mathbb{S}$ is not a whole sphere.} We may change the notation ${\cal S}\left[\mathbb{S}\right]$ to ${\cal S}(r)$, because then (\ref{51c}) defines a function of $r$:
\begin{equation}
{\cal S}(r)=\frac{1}{4G\hbar}A(r)f(r), \qquad A(r)=\int_{\mathbb{S}(r)}{\rm d}A.
\label{210}
\end{equation}

\item[{\it iii)}] Eq. (\ref{210}) now defines ${\cal S}$ as a function of $r$: we have defined a (radially symmetric) entropy function on the spacetime whose metric is (\ref{7}).

\item[{\it iv)}] Division by $4G\hbar$ renders ${\cal S}$ dimensionless and allows for a direct comparison with Bekenstein--Hawking entropy.

\item[{\it v)}] Indeed, on the spherical surface $\mathbb{S}(r)$ defined by $r={\rm const}$, the term $A(r)f(r)$ is constant. Then ${\cal S}(r)$ yields the Landauer entropy corresponding to the spherical  surface with area $A(r)$. This Landauer entropy differs from the Bekenstein--Hawking entropy of the same sphere (assuming the latter is the horizon of a black hole) by the constant numerical factor $f(r)$.

\item[{\it vi)}] The second law for the test--particle entropy proved after (\ref{seis}) implies the second law for the congruence entropy (\ref{51c}).

\item[{\it vii)}] The congruence considered must not converge, as one might otherwise end up in a singularity \cite{ELLISHAWKING}. Our choice of outward directed radial geodesics circumvents this problem, at least in all the examples considered below.

\end{itemize}

\section{Overview of results} \label{results}

The tables summarise the analysis outlined above as applied to a number of relevant metrics.

Beyond the specific examples presented here, it is always possible to arrange the definition of $S(r)$ in (\ref{9c}) in such a way that the property of monotonous increase in time, Eq. (\ref{seis}), will hold for an arbitrary metric (\ref{7}). To establish this point we first observe that the function $f(r)$ in Eq. (\ref{7}) is assumed smooth. As illustrated in the example of Schwarzschild--de Sitter space, one can always define $S(r)$ piecewise, in order to ensure that $f'(r)>0$ will hold over a certain interval $I$ of values of $r$. This interval $I$ is delimited by two consecutive roots of the equation $f'(r)=0$. Whenever $f'(r)$ changes sign, it suffices to flip the sign in the definition of  $S(r)$ in order to have $f'(r)\geq 0$, {\it i.e.}\/, ${\rm d}S/{\rm d}t \geq 0$.

\begin{table*}
    \centering
	\caption{Lapse function $f^2(r)$ for different non-rotating spacetimes in four dimensions assuming spherical symmetry.}
	\label{tab:1}       
    \begin{tabular}{c|c}
		\hline\noalign{\smallskip}  \rowcolor{olive!30}
		\textbf{spacetime} & $f^2(r)$  \\
\rowcolor{gray!30}
        \noalign{\smallskip} \hline
        \noalign{\smallskip} \hline
          \noalign{\smallskip}      
        spherical Rindler             & $2\kappa(r-r_0)$ \\
        Schwarzschild                 & $1-R/r$          \\
\rowcolor{gray!30}    
        de Sitter                     & $1-r^2/L^2$      \\
        Schwarzschild--de Sitter      &  $1-R/r-r^2/L^2$ \\
\rowcolor{gray!30}
        anti de Sitter                & $1+r^2/L^2$      \\
        Schwarzschild--anti de Sitter & $1-R/r+r^2/L^2$  \\
        \noalign{\smallskip}\hline
        \noalign{\smallskip}\hline
	\end{tabular}
\end{table*}

\begin{table*}
    \centering
	\caption{Landauer entropy of a radial test particle: $S(r)=f(r)$ (plus a sign flip in de Sitter and Schwarzschild--de Sitter, to guarantee $f'(r)>0$).}
	\label{tab:2}       
    \begin{tabular}{c|c}
		\hline\noalign{\smallskip}  \rowcolor{olive!30}
		\textbf{spacetime} & $S(r)$  \\
\rowcolor{gray!30}
        \noalign{\smallskip} \hline
        \noalign{\smallskip} \hline
          \noalign{\smallskip}      
        spherical Rindler & $\left(2\kappa(r-r_0)\right)^{1/2}$ \\
        Schwarzschild & $\left(1-R/r\right)^{1/2}$ \\
\rowcolor{gray!30}    
        de Sitter  & $-\left(1-r^2/L^2\right)^{1/2}$ \\
        Schwarzschild--de Sitter & 
$\left\{ 
\begin{array}{ll}
+\left(1-R/r-r^2/L^2\right)^{1/2}, \qquad 0 \ <r<r_2\\
-\left(1-R/r-r^2/L^2\right)^{1/2}, \qquad r_2<r<r_1
\end{array}
\right.$    \\
\rowcolor{gray!30}
        anti de Sitter & $\left(1+r^2/L^2\right)^{1/2}$    \\ 
        Schwarzschild--anti de Sitter & $\left(1-R/r+r^2/L^2\right)^{1/2}$ \\ 
        \noalign{\smallskip}\hline
        \noalign{\smallskip}\hline
	\end{tabular}
\end{table*}

\begin{table*}
    \centering
	\caption{Landauer entropy  of a radial congruence orthogonally traversing a complete sphere $\mathbb{S}^2(r)$: ${\cal S}(r)=\pi r^2 f(r)/G\hbar$.}
	\label{tab:3}       
    \begin{tabular}{c|c}
		\hline\noalign{\smallskip}  \rowcolor{olive!30}
		\textbf{spacetime} & ${\cal S}(r)$  \\
\rowcolor{gray!30}
        \noalign{\smallskip} \hline
        \noalign{\smallskip} \hline
          \noalign{\smallskip}      
        spherical Rindler & $\pi r^2\left(2\kappa(r-r_0)\right)^{1/2}/G\hbar$\\
        Schwarzschild &  $\pi r^2\left(1-R/r\right)^{1/2}/G\hbar$\\ 
\rowcolor{gray!30}    
        de Sitter  & $-\pi r^2\left(1-r^2/L^2\right)^{1/2}/G\hbar$  \\
        Schwarzschild--de Sitter &  
$\left\{
\begin{array}{ll}
+\frac{\pi r^2}{G\hbar}\left(1-R/r-r^2/L^2\right)^{1/2}, \quad r_1<r<r_2\\
-\frac{\pi r^2}{G\hbar}\left(1-R/r-r^2/L^2\right)^{1/2},\quad  r_2<r<r_3
\end{array}
\right.$   \\
\rowcolor{gray!30}
        anti de Sitter & $\pi r^2\left(1+r^2/L^2\right)^{1/2}/G\hbar$ \\
       Schwarzschild--anti de Sitter &  $\pi r^2\left(1-R/r+r^2/L^2\right)^{1/2}/G\hbar$\\
        \noalign{\smallskip}\hline
        \noalign{\smallskip}\hline
	\end{tabular}
\end{table*}

\section{Discussion}\label{Disc}

There exist several natural, but conceptually different, ways in classical and quantum gravity in which one can associate an entropy to a given spacetime metric. Which entropy is ``best” depends on what one would like the entropy to measure: horizon microstates \cite{LAWS, BEKENSTEIN2, GIBBONS1, GIBBONS2, HAWKING}, information bounds \cite{BOUSSO}, coarse graining of gravitational degrees of freedom \cite{PENROSE}, quantum entanglement \cite{RYU}. More geometric notions of entropy may involve, {\it e.g.}\/, the Weyl tensor and the Bel--Robinson tensor \cite{ELLIS, WALD}; etc.

The second law of thermodynamics is expressed through an inequality; so is Landauer's principle \cite{LANDAUER}. It has been proved  that black hole evaporation saturates the Landauer inequality \cite{CORTES}; recent works on entropic cosmology where a natural validation of the second law of thermodynamics (and Landauer's principle), including viscosity, is presented, are refs. \cite{ODINTSOV, PAUL}; black hole area quantisation by imposing Landauer's principle has been studied in ref. \cite{ANANIAS}.

In this letter we have also made use of Landauer's principle regarded as a saturated inequality, {\it i.e.}\/, as an equality. We have used this fact in order to first define a Landauer entropy $S(r)$ for a test particle in radial, geodesic motion within a static, radially symmetric gravitational field; this definition is given in  Eq. (\ref{9c}). Our definition is motivated in entropic gravity \cite{VERLINDE}, to the entropy of which our own entropy reduces in the Newtonian limit. We have explicitly computed the Landauer entropy in a number of relevant spacetimes and verified that it satisfies the second law of thermodynamics, namely, monotonous increase in time, whenever evaluated along outward directed, radial geodesics.

{}Finally we have considered congruences of radial geodesics in order to define the Landauer entropy ${\cal S}(r)$ for a whole congruence as the surface integral of  $S(r)$ for test particles. Restricting for simplicity to a sphere of radius $r$, this definition is Eq. (\ref{210}). The resulting entropy ${\cal S}(r)$ also satisfies a second law; moreover it scales with the area. 

Our analysis, summarised in tables \eqref{tab:1}, \eqref{tab:2} and \eqref{tab:3}, 
makes crucial use of the assumption of spherical symmetry, but this simplifying assumption pays off. Indeed, the Landauer entropy ${\cal S}(r)$ of a sphere with radius $r$, and the Bekenstein--Hawking entropy ${\cal S}_{\rm BH}(r)$ of a black hole whose horizon is that same sphere, are related as per ${\cal S}(r)/{\cal S}_{\rm BH}(r)=f(r)$. Now $f(r)$ is constant on a sphere of radius $r$, so the two entropies are mere scalar multiples of each other, with a proportionality constant the is determined by the metric on spacetime. Without making any assumptions concerning the microscopic degrees of freedom underlying the Bekenstein--Hawking entropy, our approach relates ${\cal S}_{\rm BH}$ to Landauer's principle. Here again we profit from the assumption of spherical symmetry.

While some notions of gravitational entropy may be covariant under spacetime diffeomorphisms, the Landauer entropy considered here is definitely not covariant. But ours is not the only instance of noninvariance under diffeomorphisms: the Rindler frame in Minkowski spacetime is a well--known example of the observer--dependence of (one concept of) entropy. Indeed this ``democracy of Rindler observers" has been advocated by Padmanabhan as fundamental to gravity \cite{PADDY1}.

\section{Appendix}

Here we present the computations pertaining to the previous tables.

\subsection{Spherical Rindler}

Assume that the metric (\ref{7}) is such that $f^2(r)$ possesses a simple zero at the horizon $r=r_0$, {\it i.e.}\/, $f^2(r_0)=0$ but $\left(f^2\right)'(r_0)=2\kappa\neq 0$. Then the near--horizon geometry of (\ref{7}) is well approximated by the spherical Rindler metric:
\begin{equation}
{\rm d}s^2=-2\kappa(r-r_0){\rm d}t^2+\frac{{\rm d}r^2}{2\kappa(r-r_0)}+r^2{\rm d}\Omega_2^2.
\label{120}
\end{equation}
The constant $2\kappa>0$ is the radial acceleration of the Rindler observer (and also proportional to the horizon temperature). Our choice $f(r)=\sqrt{2\kappa(r-r_0)}$ produces
\begin{equation}
f'(r)=\frac{\kappa}{\sqrt{2\kappa(r-r_0)}}>0
\label{121}
\end{equation}
for all $r\in(r_0,\infty)$. Therefore the Landauer entropy function for a test particle,
\begin{equation}
S_{\rm Rind}(r)=\left(2\kappa(r-r_0)\right)^{1/2},
\label{122}
\end{equation}
increases monotonically along outward directed, radial geodesics $r=r(t)$.

\subsection{Schwarzschild}

Schwarzschild spacetime has an exterior metric that, in static coordinates, reads
\begin{equation}
{\rm d}s^2=-\left(1-\frac{R}{r}\right){\rm d}t^2+\left(1-\frac{R}{r}\right)^{-1}{\rm d}r^2+r^2{\rm d}\Omega^2_{2},
\label{23}
\end{equation}
where $f^2(r)=1-R/r$ with $R$ the Schwarzschild radius, and $r\in(R,\infty)$. Now
\begin{equation}
f'(r)=\frac{1}{2}\left(1-\frac{R}{r}\right)^{-1/2}\frac{R}{r^{2}}
\label{24}
\end{equation}
and $f'(r)>0$ for all $r\in(R,\infty)$. Hence the Landauer entropy function for a test particle, 
\begin{equation}
S_{\rm Sch}(r)=\left(1-\frac{R}{r}\right)^{1/2},
\label{25}
\end{equation}
is monotonically increasing along outward directed, radial geodesics $r=r(t)$.

\subsection{De Sitter}

For de Sitter spacetime we have, in static coordinates,
\begin{equation}
{\rm d}s^2=-\left(1-\frac{r^2}{L^2}\right){\rm d}t^2+\left(1-\frac{r^2}{L^2}\right)^{-1}{\rm d}r^2+r^2{\rm d}\Omega^2_{2},
\label{16}
\end{equation}
where $L$ is the de Sitter radius; here $f^2(r)=\left(1-r^2/L^2\right)$ and $r\in(0,L)$. Since
\begin{equation}
f'(r)=-\left(1-\frac{r^2}{L^2}\right)^{-1/2}\frac{r}{L^2},
\label{73}
\end{equation}
it holds that $f'(r)<0$ for all $r\in(0,L)$. Flipping the sign we arrive at 
\begin{equation}
S_{\rm dS}(r)=-\left(1-\frac{r^2}{L^2}\right)^{1/2},
\label{18}
\end{equation}
which is monotonically increasing along outward directed, radial geodesics $r=r(t)$.

\subsection{Schwarzschild--de Sitter}

Here the metric reads
\begin{equation}
{\rm d}s^2=-\left(1-\frac{R}{r}-\frac{r^2}{L^2}\right){\rm d}t^2+\left(1-\frac{R}{r}-\frac{r^2}{L^2}\right)^{-1}{\rm d}r^2+r^2{\rm d}\Omega^2_{2},
\label{50}
\end{equation}
with $R$ and $L$ as above, and
\begin{equation}
f(r)=\left(1-\frac{R}{r}-\frac{r^2}{L^2}\right)^{1/2}.
\label{60}
\end{equation}
We will assume the parameters $R$ and $L$ so tuned that the coordinate $r$ in Eq. (\ref{50}) runs over the interval $(r_1,r_3)$, and such that
\begin{equation}
f'(r)=\frac{1}{2}\left(1-\frac{R}{r}-\frac{r^2}{L^2}\right)^{-1/2}\left(\frac{R}{r^2}-\frac{2r}{L^2}\right)
\label{61}
\end{equation}
changes from positive to negative at $r_2=(RL^2/2)^{1/3}\in(r_1,r_3)$. Therefore, for outgoing geodesics to cause an entropy increase, we need to define
\begin{equation}
S_{\rm Sch-dS}(r)=\left\{\begin{array}{ll}
\left(1-\frac{R}{r}-\frac{r^2}{L^2}\right)^{1/2}, \qquad r_1<r<r_2\\
-\left(1-\frac{R}{r}-\frac{r^2}{L^2}\right)^{1/2}, \,\quad r_2<r<r_3
\end{array}\right.
\label{64}
\end{equation}
as the entropy of a test particle.

\subsection{Anti de Sitter}

In static coordinates, anti de Sitter spacetime has the metric
\begin{equation}
{\rm d}s^2=-\left(1+\frac{r^2}{L^2}\right){\rm d}t^2+\left(1+\frac{r^2}{L^2}\right)^{-1}{\rm d}r^2+r^2{\rm d}\Omega^2_{2},
\label{20}
\end{equation}
where $r\in(0,\infty)$ and $f^2(r)=\left(1+r^2/L^2\right)$. One finds
\begin{equation}
f'(r)=\left(1+\frac{r^2}{L^2}\right)^{-1/2}\frac{r}{L^2},
\label{21}
\end{equation}
so $f'(r)>0$ for all $r\in(0,\infty)$. The Landauer entropy function for test particles,
\begin{equation}
S_{\rm AdS}(r)=\left(1+\frac{r^2}{L^2}\right)^{1/2},
\label{22}
\end{equation}
is monotonically increasing along outward directed, radial geodesics $r=r(t)$.

\subsection{Schwarzschild--anti de Sitter}

Here the metric is given by
\begin{equation}
{\rm d}s^2=-\left(1-\frac{R}{r}+\frac{r^2}{L^2}\right){\rm d}t^2+\left(1-\frac{R}{r}+\frac{r^2}{L^2}\right)^{-1}{\rm d}r^2+r^2{\rm d}\Omega^2_{2},
\label{50b}
\end{equation}
where $r\in(r_1,\infty)$ and $r_1$ is the unique (real) solution\footnote{For simplicity we assume the parameters $R$ and $L$ adjusted to have just one real solution.} of 
\begin{equation}
f(r)=\left(1-\frac{R}{r}+\frac{r^2}{L^2}\right)^{1/2}=0.
\label{70}
\end{equation}
Now
\begin{equation}
f'(r)=\frac{1}{2}\left(1-\frac{R}{r}+\frac{r^2}{L^2}\right)^{-1/2}\left(\frac{R}{r^2}+\frac{2r}{L^2}\right)>0
\label{71}
\end{equation}
for all $r\in(r_1,\infty)$. Hence the Landauer entropy function for a test particle, 
\begin{equation}
S_{\rm Sch-AdS}(r)=\left(1-\frac{R}{r}+\frac{r^2}{L^2}\right)^{1/2},
\label{72}
\end{equation}
is monotonically increasing along outward directed, radial geodesics $r=r(t)$.

\vskip0.5cm
\noindent
{\bf Acknowledgements}\\
The authors acknowledge support by project MCIN/AEI/10.13039/501100011033 and by ERDF, “A way of making Europe”, under project PID2024-162480OB-I00.

\end{document}